\documentstyle[aps]{revtex}
\begin{document}
\title{Finite Number and Finite Size Effects 
in Relativistic Bose-Einstein Condensation
}
\author{K. Shiokawa$^{1,2}$\thanks
{  E-mail address: kshiok@phys.ualberta.ca.
   Present address: 
Theoretical Physics Institute,  
University of Alberta,    
Edmonton, Alberta T6G 2J1, Canada
}
 and B. L. Hu$^{1}$\thanks
{  E-mail address: hub@physics.umd.edu.
}
\\
{\small $^1$ Department of Physics, University of Maryland, College Park,
MD 20742, USA}\\
{\small $^2$ Center for Nonlinear Studies, Hong Kong Baptist University,
Kowloon Tong, Hong Kong
}\\
\small{\it(umdpp 98-102, To appear in Physical Review D)}
}
\maketitle
\begin{abstract}
 Bose-Einstein condensation
 of a relativistic ideal Bose gas in a rectangular 
 cavity is studied.
 Finite size corrections to the critical temperature
 are obtained by the heat kernel method. 
 Using zeta-function regularization of one-loop 
 effective potential, lower dimensional critical 
 temperatures are calculated.
 In the presence of strong anisotropy, 
 the condensation is shown to occur in multisteps.
 The criteria of this behavior is that  
 critical temperatures corresponding to lower dimensional
 systems are smaller than the three dimensional critical temperature.

\end{abstract}
\section{Introduction}
%
%
Bose-Einstein condensation (BEC) 
first predicted in 1924 \cite{Bose24,Einstein24,Huang,Pathria} 
has become an explosive field of
research in recent years \cite{BEC95,AEMWC95,DMADDKK95,MADKDK96}.
Due to the temperature dependence of the chemical potential
in a Bose-Einstein distribution function,
macroscopic number of bosons start accumulating
onto the ground state at the critical temperature.
This phenomenon is typically phrased
as the phase density of particles being large enough such that
all particles characterized by 
the de Broglie wavelength overlap 
to form a condensate.
Liquid helium, which becomes a superfluid at 
the transition, has until recently been known as the only substance
which shows this behavior.
However, strong interaction between helium atoms 
has been an obstacle for a complete understanding 
of the mechanism of condensation.

Recent technological progress in
atom cooling techniques made it possible to achieve 
Bose-Einstein condensation for neutral atoms.
Weakly interacting nature of these atoms enables
one to understand the condensation within the perturbative framework.
It also provides an ideal testing ground for some fundamental aspects 
of quantum mechanics in a controlled environment.
Studies of similar phenomena,
stimulated by  rapid progress in this subject, 
are no longer restricted to    
condensed-matter physics and atom-optics but start to involve other areas
in physics, such as nuclear-particle and astrophysics \cite{BEC95}.

In this paper we study the effect of a finite size container
on the condensation.
For a finite size system, the absence of thermodynamic 
limit alters various critical behaviors  
defined and expected for a bulk system \cite{Fisher71,BarFis73,Barber83}.
Thermodynamic quantities such as the free energy 
has a surface term which vanishes in the thermodynamic limit, causing a shift
in the critical point.
Finite size effects in Bose-Einstein condensation manifest themselves
as the rounding-off of the kink in the specific heat 
at the critical temperature\footnote
{ The bulk specific heat of an ideal Bose gas is not divergent,
but shows the discontinuity in its derivative.
}.
Off-diagonal long range order stays in a finite range 
at the critical point.

For a finite size quantum system, 
the invariant operator of small fluctuations has a 
discrete spectrum. 
The relevant dimensionless parameter $\eta_i$
which characterizes finite size effects near the transition
is given by
$\eta_i = \lambda_{\theta d B} / L_i $ for a nonrelativistic system,
where $\lambda_{\theta d B}$ is the thermal de Broglie wavelength 
and $L_i (i = 1,2,3)$ are the system sizes in the three spatial 
directions.
$\eta_i = \beta / L_i $ for a relativistic system, and 
$\eta_i = \beta \omega_i$ for a harmonic oscillator with natural 
frequencies $\omega_i (i = 1,2,3)$.\footnote{
In this paper, we use the units where $k_B = \hbar = 1$, which renders 
$\eta_i$  dimensionless.  
The results in ordinary units can be obtained by
replacing $\omega$ here by $\hbar \omega$ 
and $T$ here by $k_B T$.}

Presence of anisotropy adds more variety to the critical behavior.
Suppose $\eta_i > 1$ for some $i$ (for a nonrelativistic case, 
this implies $L_i < \lambda_{\theta d B}$)
 then only the lowest mode 
in the i-th direction contributes significantly to the dynamics 
of the system.  
The motion in the i-th direction 
is frozen out and the system has an infrared behavior effectively equivalent 
to a system with one less dimension in that direction 
\cite{Krueger68,Sonin69,HuOco84,HuOco87,OSH88,StyHu89}.
Thus we can classify the dynamics with an effective infrared dimension
(EIRD) into the following four cases 
dependent on the degree of anisotropy:

Case 1; ~~$\eta_1, \eta_2, \eta_3 > 1$
$~~\rightarrow$ EIRD
 = 0,

Case 2; ~~$\eta_1, \eta_2 > 1 > \eta_3$
$\rightarrow$ EIRD = 1,
 
Case 3; ~~$\eta_1 > 1 > \eta_2, \eta_3$
$\rightarrow$ EIRD = 2,

Case 4; ~~$1 > \eta_1, \eta_2, \eta_3$
$~~\rightarrow$ EIRD = 3.  \\
\noindent
In this paper, we mainly study Case 4 where modes in all three directions
are excitable.
As the temperature is lowered, the crossover behavior between
higher- and lower-dimensional excitations
can be observed. 
Each mode is labeled by three quantum numbers associated with
the excitation energy in each direction.
In the presence of strong anisotropy,
with these quantum numbers, it is meaningful 
to split the whole particle spectum into
either zero, one, two, or three-dimensional excitations.
The ground state, being the state with the lowest quantum numbers, 
is viewed as a zero-dimensional excitation.
Let us denote the number of modes excited in the corresponding
directions as $N_0, N_1, N_2, N_3$, respectively.

We can define an n-dimensional critical temperature $T_{nD}$ ($n=1,2,3$) 
as the temperature at which 
all the n-dimensionally excited modes are saturated:
\begin{eqnarray}
\mbox{3-dimensional temperature;} ~N&=&N_3(T_{3D}),
  \label{def3d}     \\
\mbox{2-dimensional temperature;} ~N&=&N_3(T_{2D}) + N_2(T_{2D}),
  \label{def2d}     \\
\mbox{1-dimensional temperature;} 
~N&=&N_3(T_{1D}) + N_2(T_{1D}) + N_1(T_{1D}),
  \label{def1d}  
\end{eqnarray}
where the thermodynamic limit in each case is taken differently.
As we set the total number of particles $N \rightarrow \infty$, 
we obtain $T_{3D}$ by tuning $\eta_1$, $\eta_2$, and $\eta_3$ to zero,
$T_{2D}$ by tuning $\eta_2$ and $\eta_3$ to zero while $\eta_1$ fixed,
and $T_{1D}$ by tuning $\eta_3$ to zero while $\eta_1$ and $\eta_2$ fixed.
Finite size corrections necessarily modify the above definitions, since they involve 
excitations in lower dimensions.
In Section 3, we will discuss this aspect in detail.

By changing the edge lengths of a cavity or oscillator
frequencies for a magnetic trap, it is possible to
control the critical temperature and realize the 
 lower dimensional condensation.
In particular, 
when $T_{1D} < T_{2D} < T_{3D}$ holds,
condensation is expected to occur in three steps:
As the temperature is lowered,
condensation into two-dimensionally excited modes
begins at $T_{3D}$ when three-dimensionally excited states saturate.
At the two dimensional critical temperature $T_{2D}$
condensation into one-dimensionally excited modes
begins. 
The condensation onto the ground state does not 
occur until one dimensional critical temperature $T_{1D}$
is reached. 

In a finite size system, the reduced chemical potential
$ \epsilon \equiv \beta ( E_0 - \mu ) $ does not vanish.
From the expression for the ground state contribution,
we can still assume $\epsilon \sim 0$ up to order $1 / N_0$.
This condition is justified as long as $N_0$ is close to the
 total number of particles,
 or equivalently, the temperature is lower
than the critical temperature.

Although work on 
BEC in relativistic systems has a long history,
modern treatment using quantum field theory 
did not begin until 1980's
\cite{HabWel81,Kapusta81,HabWel82,BenDod91,BBD91}.
At relativistic temperatures, 
$T > 2 m$, where $m$ is the mass of the relativistic field,
pair creation-annihilation effects become nonnegligible,
and the particle number is no longer conserved.
However, the global $U(1)$ gauge symmetry of the Hamiltonian guarantees 
the existence of a conserved charge based on Noether's theorem.    
The net charge $Q$ in relativistic field theory is given by
\begin{eqnarray}
  Q = 
 \sum_{l}[
 \frac{1}{e^{\beta (E_l - \mu)} - 1 }
-
 \frac{1}{e^{\beta (E_l + \mu)} - 1 }
         ]   
	  \label{f01}
\end{eqnarray}
Particles and anti-particles have chemical potentials opposite in sign due
to the fact that they carry opposite charges.
Taking this fact into account, 
the positive-definiteness of the particle number
of the particles and antiparticles with energy $E_N$ 
requires that $| \mu | \leq m$.

Another important point is 
the relation between spontaneous symmetry breaking 
(SSB) and Bose-Einstein condensation  \cite{Kapusta81}.
Condensation into the ground state results 
in a nonzero vacuum expectation value of the field.
Hence BEC can be interpreted as a SSB of the local gauge symmetry.
This argument presumes that 
the chemical potential reaches its critical value
at the critical temperature.
However, for a finite system, this is generally not the case.


%
 Bose-Einstein condensation of a relativistic noninteracting 
 quantum field in a rectangular  cavity is studied in this paper. 
Similar aspects for a nonrelativistic ideal Bose gas 
 in anisotropic magnetic traps is discussed in a companion paper \cite{Shiokawa99}. 
 In Section 2,
we derive the effective action which includes 
one-loop quantum corrections
to the classical action and use the
$\zeta$-function regularization to evaluate the grand canonical
thermodynamic potential \cite{DowCri76,Hawking77,Camporesi90,EORBZ94}.
The generalized $\zeta$-function is written in terms of $\theta$-functions
via a Mellin transformation.
We then use the asymptotic expansion of $\theta$-functions
for a large system size and a small massive field
to see the finite size correction to the total charge
and the critical temperature. 
This asymptotic expansion is a special case of the more general
class of short time expansion of the heat kernel which is 
used for spectral analysis on an arbitrary differentiable
manifold \cite{BalBlo70}. 
The terms in the expansions correspond to the volume (Weyl),
area, and  edge contributions, etc \cite{Camporesi90,Kac66,McKSin67}.
In Section 3, 
we consider the effects of accidental degeneracy in a discrete spectrum
and show that
the highly oscillating behavior of the density of states is large enough to 
dominate over
the higher order terms in the asymptotic expansion.
We introduce an infrared cutoff to include the lowest mode contribution 
properly and estimate the lower-dimensional critical temperature accurately.
In the last part of this paper, 
we discuss the multistep behavior of condensation process in the presence of 
strong anisotropy \cite{Sonin69,DruKet97}.
The conditions for 
one, two, and three-dimensional condensations are clarified.
The relevant critical temperatures are obtained.

As discussed and observed in \cite{BPK87,EJMWC96} for a weakly interacting
gas, the corrections to bulk ideal-gas ground state occupation number and 
critical temperature are well-explained by the finite size effects. 	
Interaction effects on those quantities are negligibly small.
Hence we expect the results discussed in this paper will still hold for weakly interacting gases.
On the other hand, interaction effects are known to affect 
higher moments such as the specific heat significantly 
and considered to be essential in explaining the observed specific heat data.
Extension of our analysis to the strongly interacting case
is a nontrivial problem which deserves further careful study. 

Bose-Einstein condensation of a relativistic gas could be relevant to
cosmology in the dark matter problem \cite{Madsen92},
or for inflationary universe \cite{KhlTka97}.
Our problem is directly related to                
condensation of positronium in a cavity discussed in \cite{PlaMil94}.
Although we restrict our study to a rectangular cavity, 
similar results are expected for systems with a finite boundary.
In the presence of an irregular 
boundary the dynamics of the system can be nonintegrable or chaotic.
The implication of our result for nonintegrable systems
of this kind is of particular interest \cite{Berry81}.
The extension to curved spacetimes can also be obtained by similar methods
\cite{SinPat84,ParZha91,Toms93,SmiTom96}.

\section{Effective Action and Heat Kernel}

\subsection{One Loop Effective Action}

The action of a free complex scalar field  
\begin{eqnarray}
  S_0[\phi] &=& \int d^4 x
  [
	  \partial^{\mu} \phi^{\dag}  \partial_{\mu} \phi
        - m^2 \phi^{\dag} \phi
  ]	  \label{f300}
\end{eqnarray}
is invariant under the global $U(1)$ gauge transformation
\begin{eqnarray}
\phi(x) \rightarrow e^{i \eta} \phi(x).
	  \label{c3f1010}
\end{eqnarray}
The corresponding Noether current is 
\begin{eqnarray}
         J_{\mu}(x) = 
      i \phi^{\dag}  \partial_{\mu} \phi 
    - i  \partial_{\mu} \phi^{\dag}  \phi ,
	  \label{f302}
\end{eqnarray}
with the total charge
\begin{eqnarray}
         Q = \int_{V} d^3 x  J_{0}(x) 
           = i \int_{V} d^3 x 
           [ \phi^{\dag} \dot{\phi} 
           - \dot{\phi}^{\dag} \phi ].
	  \label{f303}
\end{eqnarray}
The integration is over the volume $V$ of the cavity, 
here assumed to be rectangular
of edge lengths $ L_i (i = 1,2,3) $. 
Decomposing $\phi(x)$ into real and imaginary parts such that 
$\phi(x) = 1 / \sqrt{2} [ \phi_1(x) + i \phi_2(x) ]$, Eq. (\ref{f300}) 
becomes
\begin{eqnarray}
  S_0[\phi] &=& \frac{1}{2} \int\!d^4 x
  [
	  \partial^{\mu} \phi_1^{\dag}  \partial_{\mu} \phi_1
        + \partial^{\mu} \phi_2^{\dag}  \partial_{\mu} \phi_2
        - {m}^2 \phi_1^{\dag} \phi_1  
        - {m}^2 \phi_2^{\dag} \phi_2 
   ].
	  \label{f304}
\end{eqnarray}

The Hamiltonian for this action is
\begin{eqnarray}
  H = \frac{1}{2} \int_{V} d^3 x
  [
	 \pi_1^{~2} + \pi_2^{~2} + (\nabla \phi_1)^2 + (\nabla \phi_2)^2
        + {m}^2 \phi_1^{\dag} \phi_1  
        + {m}^2 \phi_2^{\dag} \phi_2 
   ],
	  \label{f305}
\end{eqnarray}
where $\pi_1 = \dot{\phi_1}, \pi_2 = \dot{\phi_2}$ are 
the momentum fields
canonically conjugate to $\phi_1$ and $\phi_2$ .
The total charge becomes  
$Q = \int d^3 x (\phi_2 \pi_1 - \phi_1 \pi_2) $.

The grand canonical partition function 
for this system when brought in 
contact with a heat bath at temperature
$T = 1 / \beta$ is given by
\begin{eqnarray}
  Z  = \mbox{Tr} e^{-\beta (\hat{H} - \mu \hat{Q})}, 
	  \label{f306}
\end{eqnarray}
where $\hat{H}$ and $\hat{Q}$ are the Hamiltonian and the total charge
 operators respectively and $\mu$ is the chemical potential.
Eq. (\ref{f306}) in Hamiltonian form has a path integral representation
\begin{eqnarray}
  Z  = \int\!D \pi D \phi \exp[ 
        \int_{0}^{\beta} d\tau \int_{V} d^3 x 
           [
             i \pi_1 \dot{\phi_1} +  i \pi_2 \dot{\phi_2} 
             - H + \mu (\phi_2 \pi_1 - \phi_1 \pi_2) 
           ]
        ],
	  \label{f307}
\end{eqnarray}
where $\dot{\phi_i} = \partial_{\tau} \phi_i$.
In the spirit of (imaginary time) finite temperature field
theory, a periodic boundary condition 
is imposed on $\phi_i$, with 
$\phi_i(0,\vec{x}) = \phi_i(\beta,\vec{x})$.
We perform an integral over the momentum field and obtain
\begin{eqnarray}
  Z  = \int D \phi  ~e^{- S[\phi]} ,
	  \label{f308}
\end{eqnarray}
with the action  
\begin{eqnarray}
  S[\phi] &=& \int_{0}^{\beta} d\tau  \int_{V} d^3 x
  [
	  \frac{1}{2} ( \dot{\phi}_{1} - i \mu \phi_{2} )^2
	+ \frac{1}{2} ( \dot{\phi}_{2} + i \mu \phi_{1} )^2
 \nonumber   \\
	&+&
	  \frac{1}{2} (\nabla \phi_1)^2 + \frac{1}{2} (\nabla \phi_2)^2
	+ \frac{1}{2} {m}^2 ( \phi_{1}^2 + \phi_{2}^2 )
  ].
	  \label{f310}
\end{eqnarray}

Using the background field decomposition 
$\phi = \phi_c + \varphi$
with fluctuation $\varphi$, and
expanding the action in Eq. (\ref{f310}) 
around the classical solution $\phi_c$ which minimizes
the action 
\begin{eqnarray}
  S[\phi] &=& S[\phi_c] + \frac{1}{2} 
           \sum_{i,j=1}^{2} 
\frac{\delta^2 S}{\delta \phi_i \delta \phi_j}  
 \varphi_i \varphi_j
               + O(\varphi^3).
	  \label{f311}
\end{eqnarray}
The partition function can be written as
\begin{eqnarray}
  Z  = e^{- \Gamma[\phi_c] } = 
  e^{- S[\phi_c] } \int D \varphi ~e^{- \frac{1}{2} 
  \Lambda_{ij}[\phi_c] \varphi_i \varphi_j }, 
	  \label{f312}
\end{eqnarray}
where $\Gamma[\phi_c]$ is the effective action
and $\Lambda_{ij}[\phi_c] \equiv 
\delta^2 S[\phi_c] / \delta \phi_i \delta \phi_j $.

The functional measure $D \varphi$ is defined as
\begin{eqnarray}
  D \varphi = \prod_n \frac{d c_n}{ \sqrt{2 \pi} l }, 
	  \label{f313}
\end{eqnarray}
where $c_n$ are the coefficients of an eigenfunction expansion of $\varphi$
and $l$ is a constant with unit of length.
Then the functional integral in Eq. (\ref{f312}) can be evaluated as
\begin{eqnarray}
  \prod_n \frac{1}{ \sqrt{2 \pi} l } 
\int_{-\infty}^{\infty} d c_n e^{- \frac{1}{2} \lambda_n c_n^2} 
      = \mbox{Det} ( l^2 \Lambda_{ij}[\phi_c] )^{-1/2}.
	  \label{f314}
\end{eqnarray}

The effective action to one loop order is given by 
\cite{BirDav,DeWitt65a,DeWitt75}
\begin{equation}
	\Gamma[\phi_c] =
	  S[\phi_c] + \frac{1}{2} \log \mbox{Det} (l^2 \Lambda_{ij}[\phi_c])
	  \label{f315}.
\end{equation}
The second term in Eq. (\ref{f315}) can be split into two parts as 
\begin{equation}
          \log \mbox{Det} (l^2 \Lambda_{ij}[\phi_c]) = 
          \log \mbox{Det} (l^2 \Lambda_{+}[\phi_c]) 
+ \log \mbox{Det} (l^2 \Lambda_{-}[\phi_c]),  
	  \label{f319}
\end{equation}
%
where $  \Lambda_{\pm} = -(\partial_{\tau} \pm \mu)^2 
- \nabla^2 + m^2 $.
The eigenvalues of $  \Lambda_{\pm} $ are given by
\begin{equation}
	\lambda^{\pm}_{n,N} =
	  (\frac{2 \pi n}{ \beta} \pm i \mu )^2 + \omega_N + m^2,
	  \label{f320}
\end{equation}
where $\omega_N$ is the eigenvalue of $- \nabla^2 $.

\subsection{$\zeta$-function Regularization}

The generalized $\zeta$-function for
an elliptic differential operator $\cal O$ is defined by
\begin{equation}
	\zeta_{\cal O} (s) = \mbox{Tr} ~{\cal O}^{-s}
	= \sum_{N} \lambda_{N}^{-s},
	  \label{f316}
\end{equation}
where $\lambda_{N}$ are eigenvalues of $\cal O$.
From Eq. (\ref{f316}),  
\begin{equation}
\zeta'_{\cal O} (s) = - \sum_{N} \lambda_{N}^{-s} \log \lambda_{N}^{-s}.
	  \label{f317}
\end{equation}
Thus 
\begin{equation}
	\log \mbox{Det}
	(l^2 {\cal O}) = \zeta_{\cal O} (0) \log l^2 - \zeta'_{\cal O} (0).
	  \label{f318}
\end{equation}

Using a Mellin trasformation defined by
\begin{equation}
	\lambda^{-s} = \frac{ 1 }{ \Gamma(s) }
	  \int_{0}^{\infty} d\tau ~\tau^{s-1}
		e^{- \lambda \tau}
	  \label{f3152}
\end{equation}
we can write 
the generalized $\zeta$-function
for $\Lambda_{ij}$ as 
\begin{eqnarray}
	\zeta_{\Lambda_{\pm}} (s) 
       &=& \mbox{Tr} ~{\Lambda_{\pm}}^{-s}
        =\sum_{N} \lambda_{\pm N}^{-s}
  \nonumber   \\  &=& \frac{ 1 }{ \Gamma(s) }
	  \int_{0}^{\infty} d\tau ~\tau^{s-1}
	\sum_{N} e^{- \lambda_{\pm N} ~\tau}
  \nonumber   \\  &=& \frac{ 1 }{ \Gamma(s) }
	  \int_{0}^{\infty} d\tau ~\tau^{s-1}
    K_{\Lambda_{\pm}}(\tau),
	  \label{af316}
\end{eqnarray}
where $\lambda_{\pm N}$ and $K_{\Lambda_{\pm}}(\tau)$ 
are eigenvalues and the heat kernels for $\Lambda_{\pm}$.

Here 
\begin{eqnarray}
K_{\Lambda_{\pm}}(\tau) =
 K_{0}(\tau) =
\sum_{n=-\infty}^{\infty} \sum_{N} 
  \exp[- \tau [
     (\frac{2 \pi n}{ \beta} + i \mu )^2 + \omega_N + m^2
            ]
     ]
	  \label{f3162}
\end{eqnarray}
yielding
\begin{eqnarray}
    \zeta_{\Lambda} (s) 
  &\equiv& \zeta_{\Lambda_{+}} (s) 
  = \zeta_{\Lambda_{-}} (s) 
   \nonumber   \\ &=& \frac{ \bar{\beta}^{2s}  }{ \Gamma(s) }
	  \int_{0}^{\infty} d\tau ~\tau^{s-1}
     		K_{0}(\bar{\beta}^{2} \tau),
	  \label{f3163}
\end{eqnarray}
where $\bar{\beta} \equiv \beta / 2 \pi$ \cite{SmiTom96,Actor87} and

\begin{equation}
	K_{0}(\bar{\beta}^{2} \tau)=
	K(\tau)
	  e^{- (m^2 - \mu^2) \bar{\beta}^2 \tau}
	  \theta_{3}(\mu \bar{\beta} \tau | i \tau / \pi),
	  \label{f31}
\end{equation}
where 
$  \theta_{3}( z | \tau) = 1 + 2 \sum_{n=1}^{\infty} e^{i \pi \tau} \cos(2 n 
z) $
 is a $\theta$-function \cite{theta}. 
The heat kernel $K(\tau)$ for $- \nabla^2 $ is defined by
\begin{eqnarray}
  K(\tau) &=&
	\sum_{N} e^{- \bar{\beta}^2 \omega_N \tau}.
	  \label{f322}
\end{eqnarray}
With this, the effective action 
can be expressed in terms of $\zeta_{\Lambda}(s)$ as 
\begin{equation}
	\Gamma[\phi_c] =
	  S[\phi_c] + \zeta_{\Lambda} (0) \log l^2 - \zeta'_{\Lambda} (0).
	  \label{f3151}
\end{equation}

We first consider Neumann boundary conditions 
at the boundary of the cavity.
The corresponding eigenfunction for $- \nabla^2 $ is 
\begin{eqnarray}
  \phi_N(x) = \sqrt{\frac{2}{L_1 L_2 L_3}}
        \cos( \frac{\pi n_1 x_1}{ L_1} ) 
        \cos( \frac{\pi n_2 x_2}{ L_2} ) 
        \cos( \frac{\pi n_3 x_3}{ L_3} ) 
	  \label{f323}
\end{eqnarray}
and the eigenvalue $\omega_N$  is  
\begin{eqnarray}
  \omega_N =  
          ( \frac{\pi n_1 }{ L_1} )^2 
	+ ( \frac{\pi n_2 }{ L_2} )^2
	+ ( \frac{\pi n_3 }{ L_3} )^2, 
	  \label{f324}
\end{eqnarray}
where $n_i = 0, 1, 2, \cdots (i = 1,2,3)$.

The eigenfunction for Dirichlet boundary conditions
can be written similarly as 
\begin{eqnarray}
  \phi_N(x) = \sqrt{\frac{2}{L_1 L_2 L_3}}
        \sin( \frac{\pi n_1 x_1}{ L_1} ) 
        \sin( \frac{\pi n_2 x_2}{ L_2} ) 
        \sin( \frac{\pi n_3 x_3}{ L_3} ) 
	  \label{f3231}
\end{eqnarray}
and the eigenvalue $\omega_N$  is  
\begin{eqnarray}
  \omega_N =  
          ( \frac{\pi n_1 }{ L_1} )^2 
	+ ( \frac{\pi n_2 }{ L_2} )^2
	+ ( \frac{\pi n_3 }{ L_3} )^2, 
	  \label{f3241}
\end{eqnarray}
where $n_i = 1, 2, \cdots (i = 1,2,3)$.

\subsection{Asymptotic Expansion of the Heat Kernel}

The heat kernel for all accessible quantum states 
is given by
\begin{eqnarray}
  K(\tau) &=&
	\sum_{N} e^{- \bar{\beta}^2 \omega_N \tau}
 \nonumber   \\
    &=&
	\sum_{n_1}^{\infty} e^{- \eta_1^{~2} \pi^2 n_1^{~2} \tau}
	\sum_{n_2}^{\infty} e^{- \eta_2^{~2} \pi^2 n_2^{~2} \tau}
	\sum_{n_3}^{\infty} e^{- \eta_3^{~2} \pi^2 n_3^{~2} \tau}
  \nonumber   \\
	 &=&
	  \frac{1}{8}
		[ \theta_{3}( 0 | i  \eta_1^{~2} \pi \tau) \pm 1]
		[ \theta_{3}( 0 | i  \eta_2^{~2} \pi \tau) \pm 1]
		[ \theta_{3}( 0 | i  \eta_3^{~2} \pi \tau) \pm 1],
	  \label{f32}
\end{eqnarray}
where $ \eta_i = \bar{\beta} /  L_i  $ for $i=1,2,3$. 
Positive (negative) signs 
correspond to Neumann (Dirichlet) boundary conditions.
If we assume  $ L_i >> \bar{\beta} $  or  $ \eta_i << 1 $ for $i = 1, 2, 3$,
we can make use of the asymptotic behavior of $\theta$-function 
as $\tau \rightarrow 0$ 
\begin{equation}
 \theta_{3}( 0 | i \tau) \rightarrow    \frac{1}{ \sqrt{ \tau}  }
	  \label{f33}
\end{equation}
to obtain the following asymptotic property 
for the heat kernel
\begin{eqnarray}
	K(\tau) 
	   & \rightarrow &
	   \frac{1}{8} 
	   [   \frac{1}{ \sqrt{ \eta_1^{~2} \pi \tau}  }  \pm 1 ]
	   [   \frac{1}{ \sqrt{ \eta_2^{~2} \pi \tau}  }  \pm 1 ]
	   [   \frac{1}{ \sqrt{ \eta_3^{~2} \pi \tau}  }  \pm 1 ]
	\nonumber   \\
	   &=&
	   \frac{1}{8 ( \pi \tau )^{3/2} \eta_1 \eta_2 \eta_3  }
	   \pm
	   \frac{1}{8 \pi \tau } (
	   \frac{1}{ \eta_1 \eta_2 } + \frac{1}{ \eta_2 \eta_3 } + \frac{1}{ 
\eta_3 \eta_1 }
	    )
	   +
	   \frac{1}{8 \sqrt{\pi \tau} }  (
	   \frac{1}{ \eta_1 } + \frac{1}{ \eta_2 } + \frac{1}{ \eta_3 }
	     )
	   \pm
	   \frac{1}{8}
	 \nonumber   \\
	   &=&
            \frac{A_3}{ \bar{\beta}^3 \tau^{3/2} }
	  + \frac{A_2}{ \bar{\beta}^2 \tau }
	  + \frac{A_1}{ \bar{\beta} \tau^{1/2} }
	  + A_0,
	  \label{f35}
\end{eqnarray}
where
\begin{eqnarray}
%
%
   	     \pm A_0 &=& 1 / 8,
	\nonumber   \\
		 A_1 &=&  L_{1} + L_{2} + L_{3} / 8 \pi^{1/2},
	\nonumber   \\
	     \pm A_2 &=&  L_{1} L_{2} + L_{2} L_{3} + L_{3} L_{1} / 8 \pi,
	\nonumber   \\
		 A_3 &=&  L_{1} L_{2} L_{3} / 8 \pi^{3/2} .
	  \label{f351}
\end{eqnarray}  
Note that the leading finite size correction $A_2$ has opposite signs 
for Neumann and Dirichlet 
boundary conditions.
This fact results in the opposite shift in the critical temperature 
as we will see in Eq. (\ref{f397}), where the coefficient
$b_{3/2}$ is proportional to $A_2$.
This expansion is equivalent to the short time expansion of the
heat kernel used in spectral analysis on an arbitrary
Riemmanian manifold.
The first term in Eq. (\ref{f35}) is the Weyl term, the second term
is the boundary contribution, etc \cite{Kac66,McKSin67}.
Since the model is integrable, the heat kernel is factorized 
into contributions from each dimension.

Using the above expression in Eqs. (\ref{f3163}) and (\ref{f31}),
we obtain the generalized $\zeta$ function for $\Lambda_{ij}$
\begin{eqnarray}
	\zeta_{\Lambda}(s) &=&
	 \frac{ \bar{\beta}^{2s} }{ \Gamma(s) }
		\sum_{k = 0}^{3}  \frac{ A_{k} }{ \bar{\beta}^k }  
		\int_{0}^{\infty} d\tau ~\tau^{s - k/2 -1}
	        e^{- (m^2 - \mu^2) \bar{\beta}^2 \tau}
	  \theta_{3}(\mu \bar{\beta} \tau | i \tau / \pi)
	\nonumber   \\
	&=&
  	 \frac{ \bar{\beta}^{2s} }{ \Gamma(s) }
               \sum_{k = 0}^{3}  \frac{ A_{k} }{ \bar{\beta}^k }  
	       \int_{0}^{\infty} d\tau ~\tau^{s - k/2 -1}
		e^{- (m^2 - \mu^2) \bar{\beta}^2 \tau}
   \nonumber   \\
   & \times &	 [
	     1 + 2 \sum_{n = 1}^{\infty}  e^{- n^2 \tau}
		\sum_{q=0}^{\infty}
	  \frac{ (-1)^q ( 2 n \mu \bar{\beta} \tau)^{2q} } { (2q)! }
	 	 ].
	  \label{f36}
\end{eqnarray}
%
%
%
%
%

Let us denote the first term in Eq. (\ref{f36}) on the right hand side
of the equality as $\zeta_1(s)$,
 the second term as $\zeta_2(s)$.
Then 
\begin{eqnarray}
	\zeta_1(s) &=&
	 \frac{ \bar{\beta}^{2s} }{ \Gamma(s) }
	  \sum_{k = 0}^{3}  \frac{ A_{k} }{ \bar{\beta}^k }  
       	       [ (m^2 - \mu^2) \bar{\beta}^2 ]^{k/2-s}
		\int_{0}^{\infty} d\tau ~\tau^{s - k/2 -1}
	\nonumber   \\
	&=&
	 \sum_{k = 0}^{3} 
	 \frac{\Gamma(s-k/2) }{ \Gamma(s) }
 	  A_{k}  (m^2 - \mu^2)^{k/2-s}  
	  \label{f37}
\end{eqnarray}
and 
\begin{eqnarray}
	\zeta_2(s) &=&
		\frac{ 2 \bar{\beta}^{2s} }{ \Gamma(s) }
  	        \sum_{k = 0}^{3}  \frac{ A_{k} }{ \bar{\beta}^k }  
		\int_{0}^{\infty} d\tau ~\tau^{s - k/2 -1}
		e^{- (m^2 - \mu^2) \bar{\beta}^2 \tau}
		\sum_{n = 1}^{\infty}  e^{- n^2 \tau}
		\sum_{q=0}^{\infty}
	  \frac{ (-1)^q ( 2 n \mu \bar{\beta} \tau)^{2q} } { (2q)! }
	\nonumber   \\
	&=&
	 \frac{ 2 \bar{\beta}^{2s} }{ \Gamma(s) }
	 \sum_{k = 0}^{3}  \frac{ A_{k} }{ \bar{\beta}^k }  
	 \int_{0}^{\infty} d\tau ~\tau^{s - k/2 -1}
	\nonumber   \\
	& &
	\times
	 \sum_{p=0}^{\infty}
	 \frac{ (-1)^p (m^2 - \mu^2)^{p} \bar{\beta}^{2p} \tau^p } { p! }
	 \sum_{n = 1}^{\infty}  e^{- n^2 \tau}
 	 \sum_{q=0}^{\infty}
	  \frac{ (-1)^q ( 2 n \mu \bar{\beta} \tau)^{2q} } { (2q)! }
	\nonumber   \\
	&=&
	 \frac{ 2 \bar{\beta}^{2s} }{ \Gamma(s) }
	 \sum_{k = 0}^{3}  \frac{ A_{k} }{ \bar{\beta}^k }  
	 \sum_{p=0}^{\infty}
	 \sum_{q=0}^{\infty}
          \zeta(2s-k+2p+2q) \Gamma(s - k/2 +p +2q)  
	\nonumber   \\ 
	& \times &
		  \frac{ (-1)^{p+q}  } { p! (2q)! }
		  (m^2 - \mu^2)^{p} \bar{\beta}^{2p} ( 2 \mu \bar{\beta} )^{2q}.
	  \label{f38}
\end{eqnarray}
 We expand Eq. (\ref{f38}), assuming $ \bar{\beta} / L_i $,
 $m \bar{\beta}$ ($\mu \bar{\beta}$), and $(m^2 - \mu^2) \bar{\beta}^2$ 
 are small quantities\footnote{
 For large size ($L_i >> \bar{\beta}$) and small mass $ m << T$, we do not 
need to assume
 higher temperature. See also \cite{Kirsten91} },
 and obtain
\begin{eqnarray}
	\zeta_2(s)
	&=&
	 \frac{ 2 \bar{\beta}^{2s} }{ \Gamma(s) }
     [
	 \frac{ A_{3} }{ \bar{\beta}^3 }  \pi^{2s-7/2} \zeta(4 - 2s) \Gamma(2 - 
s)
       + \frac{ A_{2} }{ \bar{\beta}^2 }  \pi^{2s-5/2} \zeta(3 - 2s) 
\Gamma(3/2 - s)
   \nonumber   \\
      &+&  \bar{\beta}^{-1} \pi^{2s-3/2} \zeta(2 - 2s) \Gamma(1 - s)
	[ A_1 - A_3 [ m^2 + (2s-2) \mu^2]  ]
   \nonumber   \\
      &+& \zeta(2s) \Gamma(s)
	[ A_0 - A_2 [ m^2 + (2s-1) \mu^2]  ]
   \nonumber   \\
      &+&  \bar{\beta} \zeta(2s+1) \Gamma(s+1/2)
  \label{f391}
   \\
   & \times &
	[ A_3 [(m^2 - \mu^2)^2 / 2 + (s+3/2)(s+1/2) 2 \mu^4 / 3] 
        - A_1 ( m^2 + 2s \mu^2)  ] 
     ].       \nonumber
\end{eqnarray}

The one-loop effective action then has the form
\begin{eqnarray}
	 \Gamma[\phi_c]
   	 &=&
	  S[\phi_c] - \zeta'_{\Lambda}(0) + \zeta_{\Lambda}(0) \log l^2
	  \nonumber   \\
	 &=&
	  S[\phi_c]
	  \nonumber   \\
	 &+& \frac{c_3}{\bar{\beta}^{3}}
	  + \frac{c_2}{\bar{\beta}^{2}}
	  + \frac{c_1}{\bar{\beta}}
			 + c_{1/2} \log \bar{\beta}
	  + c_0 
 	  + c_{-1/2} \bar{\beta} \log \bar{\beta}  
         + \cdots,
	  \label{f392}
\end{eqnarray}
where
\begin{eqnarray}
 c_3 &=& - \frac{\pi^{1/2}}{45} A_3,   
\nonumber   \\
 c_2 &=& - \frac{\zeta(3) A_2} {\pi^2},
\nonumber   \\
 c_1 &=& - \frac{\pi^{1/2}} {3}  [ A_1 - A_3 [ m^2 - 2 \mu^2]  ],
\nonumber   \\
 c_{1/2} &=& 2  [ A_0 - A_2 [ m^2 - \mu^2]  ], 
\nonumber   \\
 c_0 &=&    2 \log(2 \pi)                                  A_0
        - \frac{\pi^{1/2}}{2} ( m^2 - \mu^2 )^{1/2}     A_1
        - [( m^2 - \mu^2 ) \log ( m^2 - \mu^2 ) 
\nonumber   \\
        &+& 2 \log(2 \pi) m^2-( 2 \log(2 \pi) +1) \mu^2]   A_2 
        - \frac{4 \sqrt{\pi} }{2}  ( m^2 - \mu^2 )^{3/2}     A_3,
\nonumber   \\
 c_{-1/2}  &=& 
        - \pi^{-1/2}   [ A_3 [ (m^2 - \mu^2)^2 + \mu^4 ] - 2 A_1 m^2 ].     
%
%
%
%
\end{eqnarray}

The total charge can likewise be written as
\begin{eqnarray}
      Q &=& 
       \frac{ -1 }{ \beta } \frac{ \partial \Gamma[\phi] }{ \partial \mu }
      \nonumber   \\
        &=&
         b_2 T^2   + b_{3/2} T \log T 
		 + b_1 T    + b_{1/2}   \log T    + b_0 + \cdots,
	  \label{f394}
\end{eqnarray}
where
\begin{eqnarray}
          b_2 &=&  \frac{ 8 \mu \pi^{3/2} }{3} A_3,      
\nonumber   \\
          b_{3/2} &=&  4 \mu    A_2,
\nonumber   \\
	  b_1 &=&      - \mu [ 4 \mu \pi^{1/2} ( m^2 - \mu^2 )^{1/2} A_3
				+ 2 [ 2 + \log ( m^2 - \mu^2 ) ]   A_2
\nonumber   \\
		&-& \frac{2 \pi^{1/2} A_1}{ ( m^2 - \mu^2 )^{1/2} }
		  - 2 \frac{A_0}{ m^2 - \mu^2 }       ],
\nonumber   \\
	  b_{1/2} &=& - \frac{\mu C}{2 \pi^{1/2}},
\nonumber   \\
	  b_0 &=& \frac{\mu}{2 \pi^{1/2}} [ 2 C (\psi(1/2) + 3 \gamma )
	 + \frac{32}{3} A_3 \mu^2 - 8 A_1 - \log (l^2) C ],
 	  \label{f3941}
\end{eqnarray}
where $ C = A_3 ( 4 \mu^2 - m^2 )  $, $\gamma$ is the Euler constant,
and $\psi$ is the Digamma function 
($\psi(1/2) = -\gamma -2 \log 2 = -1.96351\cdots$).
Note that $b_2$ gives 
the bulk term discussed in \cite{HabWel81}.

The above derivation is based on an asymptotic expansion 
of the heat kernel 
which assumes a continuum spectrum.
The same assumption may not be justified in low dimensions
where the density of states does not increase as 
rapidly with energy due to the restricted 
degrees of freedom.
Thus,
the continuum spectral approximation has to be modified
accordingly.
Indeed this type of expansion does not reproduce 
the bulk term which appeared in 
\cite{HabWel81} in the two dimensional case \cite{Toms93}.
Also it is not straightforward to define 
one dimensional critical temperature
$T_{1D}$ with the method described above 
due to the existence of inverse powers of $m^2 - \mu^2$ in Eq. (\ref{f3941}).
It is known that the chemical potential does not reach its critical 
value for a finite system at the critical temperature.
For a relativistic field theory in curved spacetime,
 this aspect has been studied
in \cite{SmiTom96}.

\section{Finite Size Effects and Multistep Condensation}

\subsection{Finite Size Effects and Discrete Spectrum}

In this section, we use an alternative method to treat the discrete spectrum
in a more appropriate way.
Rewriting the heat kernel for all accessible states in Eq. (\ref{f32}) gives
\begin{eqnarray} 
  K(\tau)   &=&
	\sum_{n_1}^{\infty} e^{- \eta_1^{~2} \pi^2 n_1^{~2} \tau}
	\sum_{n_2}^{\infty} e^{- \eta_2^{~2} \pi^2 n_2^{~2} \tau}
	\sum_{n_3}^{\infty} e^{- \eta_3^{~2} \pi^2 n_3^{~2} \tau}
  \nonumber   \\
	 &=&
	\sum_{n_1}^{\infty} 
	\sum_{n_2}^{\infty} 
	\sum_{n_3}^{\infty} 
          q_1^{n_1^{~2}}  q_2^{n_2^{~2}}  q_3^{n_3^{~2}},
 	  \label{f332}
\end{eqnarray}
where $q_i = e^{- \eta_i^{~2} \pi^2 \tau}$. 
Throughout this section, we assume that $L_3$ is an integral multiple of $L_1$ 
and $L_2$,
such that $ a_1 L_1 = a_2 L_2 = L_3 $
for some integers $a_1, a_2$.\footnote{
Integer assumption here is not essential but for calculational convenience.}
$K(\tau)$ becomes
\begin{eqnarray}
  K(\tau)  &=&
	\sum_{n_1}^{\infty} 
	\sum_{n_2}^{\infty} 
	\sum_{n_3}^{\infty} 
     q_3^{ a_1^{~2} n_1^{~2} + a_2^{~2} n_2^{~2} + n_3^{~2} }
\nonumber   \\
 	&=&
	\sum_{n = 0}^{\infty} 
     r_3(n)  q_3^{~n}, 
 	  \label{f333}
\end{eqnarray}
where $r_3(n)$ is the number of solutions of the Diophantine 
equation $ n = a_1^{~2} n_1^{~2} + a_2^{~2} n_2^{~2} + n_3^{~2}$
in natural numbers \cite{Hardy15}.

Let us define the function ${\cal N}_3(\varepsilon)$ 
which counts the number of 
points with integer coordinates
inside the ellipsoid whose x,y,z-intercept are 
$\sqrt{\epsilon} / a_1$, $\sqrt{\epsilon} / a_2$, $\sqrt{\epsilon}$, 
respectively.\footnote 
{Here we rewrite $n$ as $\varepsilon$ (dimensionless energy).}
In Appendix.A, we give a derivation of 
the exact formula for ${\cal N}_d(\varepsilon)$ for arbitrary dimension
$d$ which gives the $d$-dimensional 
cumulative density of states $\bar{\cal N}_d(\varepsilon)$ in the way
described below.

For $d=1,2,3$, they are related to each other by: 
\begin{eqnarray}
  {\cal N}_3(\varepsilon) &=& 
  8 \bar{\cal N}_3(\varepsilon)
  \pm
  12 \bar{\cal N}_2(\varepsilon)
  +
  6 \bar{\cal N}_1(\varepsilon)
  \pm
  1,
  \nonumber   \\        
   {\cal N}_2(\varepsilon) &=& 
  4 \bar{\cal N}_2(\varepsilon)
  \pm
  4 \bar{\cal N}_1(\varepsilon)
  +
  1,
  \nonumber   \\       
  {\cal N}_1(\varepsilon) &=& 
  2 \bar{\cal N}_1(\varepsilon)
  \pm
  1,
	  \label{totalcdos2}
\end{eqnarray}
where the upper (lower) signs 
correspond to Dirichlet (Neumann) boundary conditions.

Inverting (\ref{totalcdos2}),
we obtain the expression for 
$\bar{\cal N}_d(\varepsilon)~(d=1,2,3)$ in terms of ${\cal 
N}_d(\varepsilon)~(d=1,2,3)$ as
\begin{eqnarray}
  \bar{\cal N}_3(\varepsilon) &=& 
  \frac{1}{8} 
  \left[
  {\cal N}_3(\varepsilon)
  \mp
  3 {\cal N}_2(\varepsilon)
  +
  3 {\cal N}_1(\varepsilon)
  \mp
  1
  \right],
  \nonumber   \\
   \bar{\cal N}_2(\varepsilon) &=& 
  \frac{1}{4} 
  \left[
  {\cal N}_2(\varepsilon)
  \mp
  2 {\cal N}_1(\varepsilon)
  +
  1
  \right],
  \nonumber   \\
  \bar{\cal N}_1(\varepsilon) &=& 
  \frac{1}{2} 
  \left[
  {\cal N}_1(\varepsilon) 
  \mp
  1
  \right].
	  \label{totalcdos3}
\end{eqnarray}

From Eqs. (\ref{totalcdos3}) and (\ref{app:CDOS}),
we readily obtain
\begin{eqnarray}
  \bar{\cal N}_3(\varepsilon) &=&
   \frac{\pi}{6} 
   \frac{\varepsilon^{3/2}}{a_1 a_2} 
\mp 
   \frac{\pi \varepsilon}{8} 
   [  \frac{1}{a_1 a_2} + \frac{1}{a_1} + \frac{1}{a_2}  ] + 
   \Delta(\varepsilon).
 	  \label{f334}
\end{eqnarray}
The first term in Eq. (\ref{f334}) simply comes from the volume
of the ellipsoid, the second term originates from compensating 
the oversubtracted points on the three coordinate planes.
The residual term $\Delta(\varepsilon)$ includes, 
in addition to the terms corresponding to
higher order contributions in the asymptotic spectral expansion,
the error of approximating 
cubes located on the surface of the sphere by a smooth surface.
This error is ascribed to what is known as accidental degeneracies
\cite{Berry81,ItzLuc86}.
A numerical plot of $\Delta(\varepsilon)$ given in Fig.1 shows that this term 
oscillates
rapidly. 
The fitting of Sup$_{\varepsilon' < \varepsilon} \Delta(\varepsilon)$ gives 
Sup$_{\varepsilon' < \varepsilon} \Delta(\varepsilon) \sim 
\varepsilon^{\gamma}$ 
where $\gamma = 0.6$. 
Since $\gamma < 1$, the first two terms in Eq. (\ref{f334}) are still 
dominant
as long as $\varepsilon >> 1$.
However, the contribution ${\cal N}_1(\varepsilon)$ arising
from overcounting the points
on coodinate axis is proportional to $\varepsilon^{1/2}$ and
smaller than the second term \cite{GroHol95a}.
Hence the fluctuating part of the cumulative density of states 
$\Delta(\varepsilon)$
in Eq. (\ref{f334}) dominates over the contributions from 
$A_1$ and $A_0$ terms in Sec. 2.3.
Similar arguments should hold in any finite size systems
regardless of whether the system is integrable or not.

For these reasons, here we properly take into account the lowest energy
gap which carries essential information about finite size effects,
and use the continuous spectrum
approximation above the lowest excited mode.
The density of states has the following form
\begin{eqnarray}
  \rho_3(\varepsilon) &=&
   \frac{\pi}{4} 
   \frac{\varepsilon^{1/2}}{a_1 a_2} 
\mp 
   \frac{\pi }{8} 
   [ \frac{1}{a_1 a_2} + \frac{1}{a_1} + \frac{1}{a_2}  ] + \cdots.
 	  \label{f335}
\end{eqnarray}
The first term is the Weyl term, the second term is the area contribution 
from the boundary.
We can easily see these terms give the same terms in the heat kernel
Eq. (\ref{f35}) 
related by a Laplace transformation.

Next we write the heat kernel in terms of the density of states as
\begin{eqnarray}
  K(\tau)  &=& 
         K_{0} + 
        \int_{\varepsilon_1}^{\infty} \rho_3(\varepsilon) q_{3}^{\varepsilon} 
d \varepsilon, 
 	  \label{f336}
\end{eqnarray}
where $\varepsilon_1=1$ corresponds to the energy level of 
the lowest excited mode and 
$K_{0}$ is the contribution in Eq. (\ref{f333}) from the ground state.
Due to the presence of a cutoff, 
one can show that 
the total charge of all excited modes is
\begin{eqnarray}
		Q &=&
		b_2 T^2 
		+
 \frac{b_{3/2} T }{2} \log \frac{T^2}{ \tilde{m}^2 } + O(\Delta),
		  \label{fseQ2}
\end{eqnarray}
where $\tilde{m}^2 \equiv \pi^2 / L_3^{~2} + m^2 - \mu^2$ and
$O(\Delta)$ is the contribution from the residual term $\Delta(\varepsilon)$ 
and will be ignored hereafter.
Then the second term in Eq. (\ref{f394}) is replaced by
$ b_{3/2} T \log T L_3  $
for large $L_3$ close to the critical temperature.

Now we evaluate the finite size correction to the critical temperature.
The bulk critical temperature $T_c^{(0)}$ is defined by
\begin{eqnarray}
	Q = b_2 T_c^{(0) 2}    .	  \label{f395}
\end{eqnarray}
From Eq. (\ref{f394}) and the above argument, the leading correction
to the bulk critical temperature for a finite system
manifests as
\begin{eqnarray}
	Q =  b_2 T_c^{~2}   
 + b_{3/2} T_c \log T_c L_3 	 .   \label{f396}
\end{eqnarray}
From Eq. (\ref{f395}) and Eq. (\ref{f396}), we obtain the finite number 
correction
to the critical temperature for small $b_{3/2}$  
\begin{eqnarray}
	 \frac{T_c}{T_c^{(0)} }=
  1 - \frac{ b_{3/2} \log(Q L_3^{~2} / b_2) }{ 4 (b_2 Q)^{1/2} }.
		  \label{f397}
\end{eqnarray}
The correction to the condensation fraction can be easily obtained as
\begin{eqnarray}
	 \frac{Q_0}{Q}
	 &=& 1 - \frac{Q_1}{Q}
	\nonumber   \\
	 &=& 1 - (  \frac{T}{T_c^{(0)} }   )^2
	 + \frac{ b_{3/2} \log(Q L_3^{~2}/ b_2) }{ 2 (b_2 Q)^{1/2} }
	 [ (   \frac{T}{T_c^{(0)} }  )^2 -  \frac{T}{T_c^{(0)} }  ]
	 \label{f398}        \\
         &-& \frac{ b_{3/2} }{ (b_2 Q)^{1/2} }
        \frac{T}{ T_c^{(0)} } \log (  \frac{T}{T_c^{(0)} }  ).
        \nonumber       
\end{eqnarray}
In Fig. 2, we plot the condensation fraction 
of the ground state as a function of the temperature.
As mentioned in Section 2.3,
the finite size correction in Eq. (\ref{f397})
gives the opposite shift in the critical temperature
whether Neumann or Dirichlet boundary condition is used.
In the case of Dirichlet boundary condition,
as we saw in Eq.(\ref{f335}), the surface term decreases the density of states.
The smaller density of states requires the excitations with higher energy 
for the condensation criteria (\ref{def3d}) to be met and 
therefore the condensation
has to start at a higher temperature.  
These results agree with those found in \cite{BarFis73,GroHol95a}.  
It is of interest to compare the results obtained here 
with those for atoms trapped in a harmonic
oscillator potential. 
One can easily see that the boundary effect in such a potential 
is due to the Neumann boundary condition \cite{Shiokawa99}: the surface
term increases the density of states which results in the decrease
of the critical temperature from the bulk value as 
observed in \cite{EJMWC96}.

\subsection{Multistep Condensation}

%
%

\subsubsection{One-dimensional Condensation}

As mentioned in the Introduction, in the presence of strong anisotropy,
condensation can occur in multisteps.
To see one-dimensional condensation, 
we require $L_{1} = L_{2} << L_{3}$, 
equivalently, $a_1 = a_2 >> 1$ where $a_1 L_1 = a_2 L_2 = L_3$
as defined in Sec. 3.1.
In such a case, it is meaningful to split all the excited quantum states into 
one, two,
and three-dimensionally excited modes in the following way.
Hereafter we focus on Neumann boundary conditions through the rest of the 
paper.

The corresponding heat kernels for these states can be defined as 
\begin{eqnarray}
 K_1(\tau) &=&
	\sum_{n_3 = 1}^{\infty} e^{- \eta_3^{~2} \pi^2 n_3^{~2} \tau},
	\nonumber   \\
 K_2(\tau) &=&
	2 \sum_{n_2 = 1}^{\infty} e^{- \eta_2^{~2} \pi^2 n_2^{~2} \tau}
	  \sum_{n_3 = 0}^{\infty} e^{- \eta_3^{~2} \pi^2 n_3^{~2} \tau},
 	\nonumber   \\
 K_3(\tau) &=&
	\sum_{n_1 = 1}^{\infty} e^{- \eta_1^{~2} \pi^2 n_1^{~2} \tau}
	\sum_{n_2 = 1}^{\infty} e^{- \eta_2^{~2} \pi^2 n_2^{~2} \tau}
	\sum_{n_3 = 0}^{\infty} e^{- \eta_3^{~2} \pi^2 n_3^{~2} \tau},
	  \label{1dcond}
\end{eqnarray}
respectively. 
The factor $2$ in $K_2(\tau)$ is due to the symmetry between
$L_1-L_3$ plane and $L_2-L_3$ plane.

Following the similar steps from Eqs. (\ref{totalcdos2}) to (\ref{f335}), 
we obtain the expression for the three-dimensional density of states as
\begin{eqnarray}
  \rho_3(\varepsilon) &=&
   \frac{\pi}{4} 
   \frac{\varepsilon^{1/2}}{a_1^2} 
-
   \frac{\pi }{24} 
   [ \frac{1}{a_1^2} + \frac{2}{a_1} ] + \cdots.
 	  \label{1dcond-DOS}
\end{eqnarray}
And the three-dimensional heat kernel in terms of
 the density of states is given as
\begin{eqnarray}
  K_3(\tau)  &=& 
      \int_{\varepsilon_1}^{\infty} \rho_3(\varepsilon) q_{3}^{\varepsilon} d 
\varepsilon, 
 	  \label{1dcond-HK}
\end{eqnarray}
where $\varepsilon_1=a_1^2$.
This gives the total charge of three-dimensionally excited modes 
\begin{eqnarray}
		Q_3 &=&
		b_2 T^{~2}
   - \frac{b_{3/2} T }{6} \log \frac{T^2}{ \tilde{m}^2 },
		  \label{1dcondQ3}
\end{eqnarray}
where $\tilde{m}^2 \equiv \pi^2 \varepsilon_1 / L_3^{~2} + m^2 - \mu^2$.

For the two dimensional heat kernel, 
\begin{eqnarray}
K_2(\tau) &=&
     2  \sum_{n_2 = 1}^{\infty} e^{- \eta_2^{~2} \pi^2 n_2^{~2} \tau}
	\sum_{n_3 = 0}^{\infty} e^{- \eta_3^{~2} \pi^2 n_3^{~2} \tau}
  \nonumber   \\
	&=&
     2  \sum_{n_2 = 1}^{\infty}
	\sum_{n_3 = 0}^{\infty} q_{3}^{a_1^{~2} n_2^{~2} + n_3^{~2}}
	=
	\sum_{n = 0}^{\infty}
     2  r_{2}(n) q_{3}^n,
		\label{1dcond-2dHK}
\end{eqnarray}
where $r_{2}(n)$ is the number of
 solutions of $n = a_{1}^{~2} n_2^{~2} + n_3^{~2}$ in natural numbers.
We write Eq. (\ref{1dcond-2dHK}) in an integral form using the density of 
states
$\rho_2 (\varepsilon) = \pi/ 2 a_1$ such that
\begin{eqnarray}
 K_2(\tau) &=&
	\int_{\varepsilon_1}^{\infty} d \varepsilon \rho_2(\varepsilon) 
q_{3}^{\varepsilon}
	\nonumber   \\
	 &=&
	  \frac{ q_{3}^{\varepsilon_1}  }{ 2 \pi a_1 \eta_3^{~2} \tau  },
		\label{1dcond-2dHK2}
\end{eqnarray}
where $\varepsilon_1 = a_1^2$ for the present choice of units.
Then from the expression of the zeta function in terms of 
the heat kernel in Eqs. (\ref{f3163}) and (\ref{f31}) we obtain
\begin{equation}
	\zeta_{\Lambda}(s) = \frac{ L_2 L_3 }{ 4 \pi }
				 \frac{ \bar{\beta}^{2s-2} }{ \Gamma(s) }
	  \int_{0}^{\infty} d\tau ~\tau^{s-2}
	  e^{- \tilde{m}^2 \bar{\beta}^2 \tau}
	  \theta_{3}(\mu \bar{\beta} \tau | i \tau / \pi),
	  \label{1dcond-zeta} 
\end{equation}
where $\tilde{m}^2$ is the same as in Eq. (\ref{1dcondQ3}).
The total charge of two-dimensionally excited states is
\begin{eqnarray}
		Q_2 &=&
 \frac{2 \mu L_2 L_3 T }{\pi} \log \frac{T^2}{ \tilde{m}^2 }. 
		  \label{1dcond-Q2}
\end{eqnarray}

For the one dimensional case,
we have
\begin{equation}
	\zeta_{\Lambda}(s) = \frac{ \bar{\beta}^{2s} }{ \Gamma(s) }
	  \int_{0}^{\infty} d\tau ~\tau^{s-1}
	  \sum_{n_3 = 1}^{\infty}
	  q_{3}^{n_3^{~2}}
	  e^{- (m^2 - \mu^2) \bar{\beta}^2 \tau}
	  \theta_{3}(\mu \bar{\beta} \tau | i \tau / \pi).
	  \label{1dcond-zeta1d}
\end{equation}
The total charge carried by one-dimensionally excited states has the form
\begin{eqnarray}
		Q_1 &=&
		\frac{2 \mu L_3^{~2} T }{\pi} \log (2 \pi).
		 \label{1dcond-Q1}
\end{eqnarray}

The three-dimensional critical temperature is reached when all the three-
dimensionally
excited modes are saturated, namely,
sum of all the modes with energy larger than $\varepsilon = a_1^2$ is equal 
to 
the total charge $Q$.
We write this condition corresponding to Eq. (\ref{def3d}) as
\begin{eqnarray}
Q = Q_3(T_{3D}).
		 \label{1dcond-T31}
\end{eqnarray}
Thus we obtain 
\begin{eqnarray}
	Q =  b_2 T_{3D}^{~2} 
	  - \frac{b_{3/2} T_{3D} }{3} \log \frac{T_{3D} L_1 }{\pi} 
	  + \frac{4 m L_2 L_3 T_{3D} }{\pi} \log \frac{T_{3D} L_1 }{\pi},
	     \label{1dcond-T32}
\end{eqnarray}
where we set $\mu=m$.
The third term is the contribution from 
two-dimensionally excited modes with energy larger than $\varepsilon = 
a_1^2$. 
We will ignore the contribution from $Q_1$ based on the argument in Sec. 3.1 
that
this term is dominated by the residual term $\Delta(\varepsilon)$.

For one-dimensional condensation to be observable, we must have
\begin{eqnarray}
T_{1D} < T_{3D}.
		 \label{1dcond-condition}
\end{eqnarray}
Furthermore for sufficiently large $L_3 >> L_1, L_2$,
comparing the expression for $Q_1$ in Eq. (\ref{1dcond-Q1}) with 
those for $Q_2$ and $Q_3$ 
in Eqs. (\ref{1dcond-Q2}) and (\ref{1dcondQ3}), we obtain
\begin{eqnarray}
Q_2(T_{1D}), Q_3(T_{1D}) << Q_1(T_{1D}).
		 \label{1dcond-Qcondition}
\end{eqnarray}
Thus we obtain $T_{1D}$ as the temperature at which one-dimensionally 
excited states saturate, i.e.
\begin{eqnarray}
Q = Q_1(T_{1D}).
		 \label{1dcond-QQ1}
\end{eqnarray}
This gives
$ T_{1D} = \pi Q [ 2 m L_3^{~2} \log (2 \pi) ]^{-1}$.

%
%

\subsubsection{Two-dimensional Condensation}

For two-dimensional condensation, 
we assume $L_{1} <<  L_{2} = L_{3}$, whence 
we split the excited quantum states into 
\begin{eqnarray}
 K_1(\tau) &=&
	2 \sum_{n_3 = 1}^{\infty} e^{- \eta_3^{~2} \pi^2 n_3^{~2} \tau}
	\nonumber   \\
 K_2(\tau) &=&
	\sum_{n_2 = 1}^{\infty} e^{- \eta_2^{~2} \pi^2 n_2^{~2} \tau}
	  \sum_{n_3 = 1}^{\infty} e^{- \eta_3^{~2} \pi^2 n_3^{~2} \tau}
 	\nonumber   \\
 K_3(\tau) &=&
	\sum_{n_1 = 1}^{\infty} e^{- \eta_1^{~2} \pi^2 n_1^{~2} \tau}
	\sum_{n_2 = 0}^{\infty} e^{- \eta_2^{~2} \pi^2 n_2^{~2} \tau}
	\sum_{n_3 = 0}^{\infty} e^{- \eta_3^{~2} \pi^2 n_3^{~2} \tau}.
	  \label{2dcond}
\end{eqnarray}
The integer $a_1 >> 1$ defined by $a_1 L_1 = L_2 = L_3$ will be used.
The factor $2$ in $K_1(\tau)$ acounts for the symmetry between
$L_2$ direction and $L_3$ direction.

The three-dimensional density of states becomes
\begin{eqnarray}
  \rho_3(\varepsilon) &=&
   \frac{\pi}{4} 
   \frac{\varepsilon^{1/2}}{a_1^2} 
+
   \frac{\pi }{24} 
   [ \frac{2}{a_1} + 1 ] + \cdots.
 	  \label{2dcond-DOS}
\end{eqnarray}
%
%
%
%
This gives the total charge carried by three-dimensionally excited modes 
\begin{eqnarray}
		Q_3 &=&
		b_2 T^{~2}
   + \frac{b_{3/2} T }{6} \log \frac{T^2}{ \tilde{m}^2 },
		  \label{2dcondQ3}
\end{eqnarray}
where $\tilde{m}^2 \equiv \pi^2 \varepsilon_1 / L_3^{~2} + m^2 - \mu^2$
and $\varepsilon_1=a_1^2$.

The two dimensional heat kernel is given by the density of states
$\rho_2 (\varepsilon) = \pi/ 4$ as
\begin{eqnarray}
 K_2(\tau) =
	\int_{\varepsilon_1}^{\infty} d \varepsilon \rho_2(\varepsilon) 
q_{3}^{\varepsilon}
           =
 \frac{ q_{3}^{\varepsilon_1}  }{ 4 \pi \eta_3^{~2} \tau  },
		\label{2dcond-2dHK}
\end{eqnarray}
where $\varepsilon_1 = 1$.
The total charge of two-dimensionally excited states is
\begin{eqnarray}
		Q_2 &=&
 \frac{\mu L_2 L_3 T }{2 \pi} \log \frac{T^2}{ \tilde{m}^2 }. 
		  \label{2dcond-Q2}
\end{eqnarray}
The total charge carried by one-dimensionally excited states is given by
$Q_1 = 4 \mu L_3^{~2} T \log (2 \pi) / \pi$.

The three-dimensional critical temperature is reached under the same 
condition
as in one-dimensional condensation,
$ Q = Q_3(T_{3D}).
		 \label{2dcond-T31}
$ 
Thus we obtain
\begin{eqnarray}
	Q =  b_2 T_{3D}^{~2} 
	 + \frac{b_{3/2} T_{3D} }{3} \log \frac{T_{3D} L_1 }{\pi} 
	 + \frac{m L_3^{~2} T_{3D} }{\pi} \log \frac{T_{3D} L_1}{\pi},
	     \label{2dcond-T32}
\end{eqnarray}
where the third term is the contribution from the two-dimensional 
modes as in Eq. (\ref{1dcond-T32}).

For two-dimensional condensation to be observable, we must have
\begin{eqnarray}
T_{2D} < T_{3D}.
		 \label{2dcond-condition}
\end{eqnarray}
Furthermore, for sufficiently large $L_2, L_3 >> L_1$,
by comparing the expression in Eqs. (\ref{2dcondQ3}) 
and (\ref{2dcond-Q2}), we have
\begin{eqnarray}
Q_3(T_{2D}) << Q_2(T_{2D}).
		 \label{2dcond-Qcondition}
\end{eqnarray}
Thus we obtain $T_{2D}$ as the temperature 
at which two-dimensionally excited states saturate,
\begin{eqnarray}
Q = Q_2(T_{2D}).
		 \label{2dcond-QQ2}
\end{eqnarray}
Then 
$ T_{2D} = Q [ \tilde{b}_{3/2} \log ( Q L_3/\tilde{b}_{3/2} ) ]^{-1} $ 
for large $Q$, where $ \tilde{b}_{3/2} = m L_2 L_3 / \pi = m L_3^2 / \pi$.

%
%

\subsubsection{Three-step Condensation}

To show three-step condensation, 
we assume $L_{1} << L_{2} << L_{3}$, or equivalently, $a_1 >> a_2 >> 1$.
The corresponding heat kernels for these states can be defined as 
\begin{eqnarray}
 K_1(\tau) &=&
	\sum_{n_3 = 1}^{\infty} e^{- \eta_3^{~2} \pi^2 n_3^{~2} \tau}
			=
	\frac{1}{2} [ \theta_{3}( 0 | i  \eta_3^{~2} \pi \tau) - 1],
	\nonumber   \\
 K_2(\tau) &=&
	\sum_{n_2 = 1}^{\infty} e^{- \eta_2^{~2} \pi^2 n_2^{~2} \tau}
	\sum_{n_3 = 0}^{\infty} e^{- \eta_3^{~2} \pi^2 n_3^{~2} \tau}
         =
	\frac{1}{4}
	[ \theta_{3}( 0 | i  \eta_2^{~2} \pi \tau) - 1]
	[ \theta_{3}( 0 | i  \eta_3^{~2} \pi \tau) + 1],
	\nonumber   \\
 K_3(\tau) &=&
	\sum_{n_1 = 1}^{\infty} e^{- \eta_1^{~2} \pi^2 n_1^{~2} \tau}
	\sum_{n_2 = 0}^{\infty} e^{- \eta_2^{~2} \pi^2 n_2^{~2} \tau}
	\sum_{n_3 = 0}^{\infty} e^{- \eta_3^{~2} \pi^2 n_3^{~2} \tau}
\nonumber   \\
			&=&
	\frac{1}{8}
	[ \theta_{3}( 0 | i  \eta_1^{~2} \pi \tau) - 1]
	[ \theta_{3}( 0 | i  \eta_2^{~2} \pi \tau) + 1]
	[ \theta_{3}( 0 | i  \eta_3^{~2} \pi \tau) + 1].
	  \label{f352}
\end{eqnarray}
The asymptotic behavior when $ 1 >> \eta_1 >> \eta_2 >> \eta_3 $ 
can be derived in a similar way as in Eq. (\ref{f35}).

%
%
%
%
%

The three-dimensional density of states becomes
\begin{eqnarray}
  \rho_3(\varepsilon) &=&
   \frac{\pi}{4} 
   \frac{\varepsilon^{1/2}}{a_1 a_2} 
+
   \frac{\pi }{24} 
   [ \frac{1}{a_1 a_2} + \frac{1}{a_1} + \frac{1}{a_2} ] + \cdots
 	  \label{3dcond-DOS}
\end{eqnarray}
%
%
%
%
and the total charge of three-dimensionally excited modes 
%
%
%
is given by the same form as in Eq. (\ref{2dcondQ3}).

From the two dimensional heat kernel, we obtain the two-dimensional density 
of states
$\rho_2 (\varepsilon) = \pi/ 4 a_2$ such that
\begin{eqnarray}
 K_2(\tau) &=&
	\int_{\varepsilon_1}^{\infty} d \varepsilon \rho_2(\varepsilon) 
q_{3}^{\varepsilon}
	\nonumber   \\
	 &=&
	  \frac{ q_{3}^{\varepsilon_1}  }{ 4 \pi a_2 \eta_3^{~2} \tau  },
		\label{f3001}
\end{eqnarray}
where $\varepsilon_1 = a_2^2$ for the present case.
This will give us the total charge of two-dimensionally excited states as
\begin{eqnarray}
		Q_2 &=&
 \frac{\mu L_2 L_3 T }{2 \pi} \log \frac{T^2}{ \tilde{m}^2 }. 
		  \label{af3002}
\end{eqnarray}
%
%
%
%
The total charge carried by one-dimensionally excited states 
has the same form as 
in Eq. (\ref{1dcond-Q1}).
%
%

The three-dimensional critical temperature is obtained as
\begin{eqnarray}
	Q =  b_2 T_{3D}^{~2} 
	 + \frac{b_{3/2} T_{3D} }{3} \log \frac{T_{3D} L_1 }{\pi} 
	 + \frac{m L_2 L_3 T_{3D} }{\pi} \log \frac{T_{3D} L_1 }{\pi}.
	 \label{3dcond-T32}
\end{eqnarray}

To observe three-step condensation, we must have
\begin{eqnarray}
T_{1D} < T_{2D} < T_{3D}.
		 \label{3dcond-condition1}
\end{eqnarray}
In addition, for large anisotropy $L_{1} << L_{2} << L_{3}$,
comparison of explicit formulas for $Q_1, Q_2$ and $Q_3$ in
Eqs. (\ref{1dcond-Q1}), (\ref{af3002}) and (\ref{2dcondQ3}) gives
\begin{eqnarray}
Q_3(T_{2D}) << Q_2(T_{2D})
		 \label{3dcond-condition2}
\end{eqnarray}
and
\begin{eqnarray}
Q_3(T_{1D}) << Q_2(T_{1D}) << Q_1(T_{1D}),
		 \label{3dcond-condition3}
\end{eqnarray}
where $T_{2D}$ ($T_{1D})$
is obtained by the saturation of the two (one)-dimensionally 
excited states as
$Q = Q_2(T_{2D})$ and $Q = Q_1(T_{1D})$.
$T_{2D}$ is given in the leading
order by
\begin{eqnarray}
	Q =   \tilde{b}_{3/2} T_{2D} \log (L_2 T_{2D}) ,	  \label{f3992}
\end{eqnarray}
where $\tilde{b}_{3/2} = m L_2 L_3 / \pi $.
We thus obtain 
$ T_{2D} = Q [ \tilde{b}_{3/2} \log ( Q L_2/ \tilde{b}_{3/2} ) ]^{-1} $ 
for large $Q$.
$ T_{1D} $ has the same form as in the one-dimensional condensation case.

From Eqs. (\ref{3dcond-condition1}) to (\ref{3dcond-condition3}),
we obtain the following inequalities 
for anisotropy parameters and the total charge corresponding to
 (A)  $T_{1D} < T_{2D}$,
 (B)  $T_{2D} < T_{3D}$, and
 (C)  $T_{1D} < T_{3D}$ as
\begin{eqnarray}
\frac{L_3}{L_2}  & >> & 
\frac{ \log (\tilde{Q}) }
     { 2 \log(2 \pi) },
    \label{simplified1} 
  \\
\frac{L_3}{L_1} & >> &
\frac{\pi \tilde{Q}}{3 [ \log \tilde{Q} ]^2}, 
    \label{simplified2}
  \\
\frac{L_3}{L_1} \left[ \frac{L_3}{L_2} \right]^2 & >> &
\frac{\pi \tilde{Q}}{12 [ \log (2 \pi) ]^2}, 
    \label{simplified3}
\end{eqnarray}
where $\tilde{Q} \equiv \pi Q / m L_2$.
In Fig. 3, 
different multistep behaviors corresponding to various ranges
of anisotropy parameters are shown.
Three-step BEC can be seen in a wide region of parameter space.
For an extremely strong anisotropy, dynamics along the short edge length
will freeze out (EIRD $<$ 3) before BEC into $Q_2$ or $Q_1$ sets in.
In such a case, we only observe two or one-step BEC. 
For a quasi-linear cavity along the vertical axis,
three-step BEC can still be observed, while for a quasi-planar cavity
along the horizontal axes, only up to two-step BEC can happen.

In Fig. 4,
the condensation fractions
$ Q_0 / Q,  Q_1 / Q, Q_2 / Q,  Q_3 / Q$ as a function of the
temperature are plotted.
In the isotropic case (Fig. 4a), 
condensation is only into the ground state.
Due to the finite size effects, condensation occurs before the critical 
temperature
is reached.
In strongly anisotropic cases, condensation occurs in steps.  
In Fig. 4b, one-dimensional condensation is seen.
$T_{3D}$ determines the onset of condensation into one-dimensionally excited 
states. 
Note that at $T_{3D}$, the ground state fraction is negligibly small.
Condensation into the ground state occurs at a much lower temperature.
In Fig. 4c, two-dimensional condensation is plotted. 
At $T_{3D}$, two-dimensional condensation manifests itself.
The critical condition $\mu = m$ is satisfied well in Fig. 4a-c.
In Fig. 4d, three-step condensation is shown. 
Three-dimensionally excited modes dominant in higher temperature
are condensed into two, one, and the ground state as the temperature is
lowered.
The deviation of $T_{3D}$ and the onset of two-dimensional components
reflects the fact that the condition $\mu = m$ is
not satisfied for the parameters chosen in Fig. 4d at $T_{3D}$.  
This is another manifestation of finite size effects and 
the result should improve near the thermodynamic limit
($\eta_i \rightarrow 0, Q \rightarrow \infty$).
The similarity between each condensation process becomes
evident in the logarithmic $T$ scale as can be seen in Fig. 4e. 
In conclusion, finite size effects on the Bose-Einstein condensation of 
a charged scalar field can lead to the multistep condensation
in the presence of strong anisotropy.

In this paper, we started from calculating the effective action to one-loop 
order using zeta function regularization.
Large volume and small mass conditions are assumed to facilitate 
an asymptotic expansion of the heat kernel 
and finite size corrections corresponding to the surface term, corner term, 
etc. are obtained.
We proceeded beyond the continuum spectrum approximation and
showed that the higher order terms in the standard asymptotic expansion are 
dominated by the contribution from the fluctuating part of the 
density of states due to accidental degeneracy.
The lowest energy gap is shown to play the crucial role 
in determining the critical temperatures for one and two-dimensional
systems.
The corresponding low-dimensional critical temperatures are calculated.
The energy spectrum and the associated heat kernel can be partitioned 
into parcels of eigenmodes excitable in dimensions 3, 2, 1, or 0. 
As the temperature is lowered, 
modes in different parcels behave quite differently in the presence of strong 
anisotropy.
When $T_{1D},T_{2D} < T_{3D}$ are satisfied, condensation occurs first into 
the lower dimensionally 
excited states at $T_{3D}$ following the ground state condensation at lower-
dimensional critical temperature.
Experimental observation of these phenomena 
can in principle be realized in an anisotropic harmonic potential
traps.
\vspace{0.3cm}

\noindent {\bf Acknowledgement}
We thank Prof. J. Weiner for helpful comments of experimental relevance and
Dr. K. Kirsten for useful references. 
K. S. appreciated the hospitality of the Center for Nonlinear Studies
at the Hong Kong Baptist University
during his visit from March to September 1998.
This work is supported in part by the U S National Science Foundation
under grants PHY94-21849.

%
%
%

\appendix
\renewcommand{\theequation}{\thesection\arabic{equation}}

\section{Cumulative Density of States}
\label{app:Cumulative Density of States}
\setcounter{equation}{0}

In this appendix, 
the exact formula for ${\cal N}_d(\varepsilon)$ for arbitrary dimension
$d$ is derived.
$d$-dimensional 
cumulative density of states $\bar{\cal N}_d(\varepsilon)$ can be obtained 
straightforwardly from ${\cal N}_d(\varepsilon)$ as we showed in Section 3.
Suppose ${\cal N}_d(\varepsilon)$ counts the number of integer solutions 
$\vec{n}$
of the equation $\varepsilon = a_1^{~2} n_1^{~2} + \cdots + a_d^{~2} 
n_d^{~2}$
where $\vec{a} \equiv (a_1, \cdots, a_d)$ is a constant $d$-dimensional vector with integer 
coordinates.
Then ${\cal N}_d(\varepsilon)$ can be written as
\begin{eqnarray}
  {\cal N}_d(\varepsilon) 
  &=& 
  \sum_{\vec{n}} \theta(\varepsilon - |\vec{a}\vec{n}|^2)
  \nonumber   \\
  &=& 
   \sum_{\vec{n}} \int d \vec{u}~ \theta(\varepsilon - |\vec{u}|^2)
   \delta^d(\vec{u} - \vec{a}\vec{n} )
    \nonumber   \\
  &=& 
  A^{-1} \sum_{\vec{l}} 
  \int d \vec{u}~ \theta(\varepsilon - |\vec{u}|^2) 
  e^{ 2 \pi i \vec{u} \cdot (\vec{l}/\vec{a})   }
    \nonumber   \\
  &=& 
   \sum_{\vec{l}} C(\vec{l}),
 	  \label{totalcdos}
\end{eqnarray}
where $A \equiv a_1 \cdots a_d$, 
$\vec{a}\vec{n} \equiv (a_1 n_1, \cdots, a_d n_d)$,
and $\vec{l} / \vec{a} \equiv ( l_1 / a_1, \cdots, l_d / a_d) $.
Summation is over all $d$-dimensional vectors with interger coordinates.
Poisson's summation formula is used to obtain the third line and
\begin{eqnarray}
C(\vec{l}) 
  &\equiv& 
  A^{-1} \int d \vec{u}~ \theta(\varepsilon - |\vec{u}|^2) 
   e^{2 \pi i \vec{u} \cdot (\vec{l}/\vec{a}) }
    \nonumber   \\
  &=& 
   A^{-1} \int d \vec{u}~ \theta(\varepsilon - |\vec{u}|^2) 
   e^{2 \pi i |\vec{u}||\vec{l}/\vec{a}| \cos \theta},
 	  \label{Fourier1}
\end{eqnarray}
where $\theta$ is the angle between $\vec{u}$ and $\vec{l}/\vec{a}$. 
We make an orthogonal coordinate transformation from $\vec{u}$ to $\vec{v}$
such that $v_1 = |\vec{u}|\cos \theta$ and write Eq. (\ref{Fourier1}) as
\begin{eqnarray}
C(\vec{l}) 
  &=& 
  A^{-1} \int d \vec{v}~ \theta(\varepsilon - |\vec{v}|^2) 
  e^{2 \pi i v_1 |\vec{l}/\vec{a}|}
  \nonumber   \\
   &=& 
  A^{-1}\frac{ \pi^{(d-1)/2} {\varepsilon}^{(d-1)/2}}{ \Gamma((d+1)/2) }
  \int_{-1}^{1} d v_1 (1 - v_1^{~2})^{(d-1)/2} e^{2 \pi i v_1 
|\vec{l}/\vec{a}|}.
	  \label{Fourier2}
\end{eqnarray}
This yields
\begin{eqnarray}
C(\vec{l})
  = 
  \left\{
  \begin{array}{ll}
  A^{-1}  
   {\varepsilon}^{d/4} 
  J_{d/2} ( 2 \pi |\vec{l}/\vec{a}| \sqrt{\varepsilon} ) /
  |\vec{l}/\vec{a}|^{d/2}     
             \hspace{1cm}   & \mbox{for} ~~\vec{l} \neq  0
     \nonumber   \\                     
  A^{-1}  {\varepsilon}^{d/4}  V_{d-1} 
                            & \mbox{for} ~~\vec{l} = 0,
  \end{array}
  \right.
  	  \label{Fourier3}
\end{eqnarray}
where 
$J_{d} ( x )$ is the Bessel function and
$V_{d} = \pi^{d/2} / \Gamma(d/2 +1)$ 
is the volume of a $d$-dimensional sphere with unit radius.

Hence we obtain
\begin{eqnarray}
  {\cal N}_d(\varepsilon) 
  &=& 
  A^{-1} {\varepsilon}^{d/4}  V_{d-1} 
  + 
  A^{-1} {\varepsilon}^{d/4} 
  \sum_{\vec{l}} 
   \frac{  J_{d/2} ( 2 \pi |\vec{l}/\vec{a}| \sqrt{\varepsilon} ) }
  { |\vec{l}/\vec{a}|^{d/2} }.
  \label{app:CDOS}
\end{eqnarray}

\newpage
\begin{flushleft}
{\large\bf Figure Captions \\}
\vspace{1cm}
\end{flushleft}

\noindent {\bf Figure 1}
The residual term 
$\Delta(\varepsilon)$ in the cumulative density of states (Eq. (\ref{f334}))
is plotted for the Neumann boundary condition.
This term shows the highly oscillating behavior
due to accidental degeneracies.  
Fig.1a is an isotropic case ($a_1 = a_2 = 1$) and 
Fig.1b is an anisotropic case ($a_1 = 10$ and $a_2 = 3$).
Supremum (dashed lines) of both curves show that the rate of increase 
is proportional to $\varepsilon^{\gamma}$ where $\gamma = 0.6$.

\vspace{0.5cm}

\noindent {\bf Figure 2}
 The condensation fraction $ Q_0 / Q $ for the 
 Neumann boundary condition
 is plotted as a function of the temperature.
$L_1=1$, $L_2=10$, $L_3=100$, $Q=10000$, and $m=0.1$. 
Dotted curve shows the bulk contribution. 
Solid curve includes the finite size correction based on Eq. (\ref{f398}).
$T_c$ denotes the finite size corrected three-dimensional critical 
temperature
defined in Eq. (\ref{f397}).

\vspace{0.5cm}
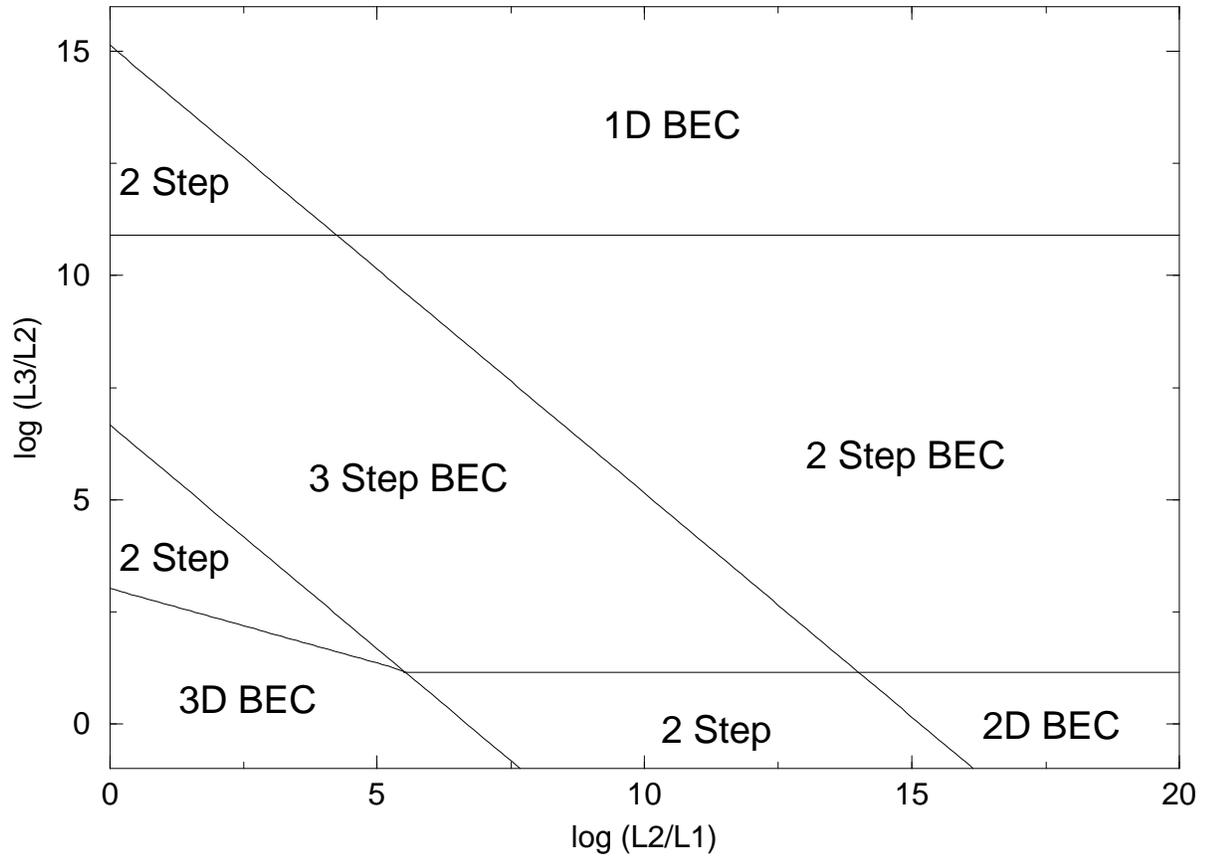
\noindent {\bf Figure 3}
Different multistep behaviors corresponding to different anisotropy parameters
$L_2/L_1$ and $L_3/L_2$ 
are indicated. The logarithmic scale is used for both axis.
$\tilde{Q} \equiv \pi Q / m L_2 = 10000$ is fixed.
Multistep BEC can only happen in an intermediate yet strongly anisotropic
regime.

\vspace{0.5cm}

\noindent {\bf Figure 4}
The condensation fractions
$Q_0 / Q$ (solid curve),  $Q_1 / Q$ (dashed curve),
 $Q_2 / Q$ (dot-dashed curve),  $Q_3 / Q$ (dotted curve)
  as a function of the
temperature are plotted for the Neumann boundary condition.
Isotropic case ($L_1=L_2=L_3=3$, $Q=100$, and $m=2$) are plotted in Fig. 4a.
Condensation is only into the ground state.
$T_c = 1.97$ is the critical temperature in Eq. (\ref{f396}).
In Fig. 4b-e, anisotropic cases are shown.
In Fig. 4b,
$L_1=2$, $L_2=2$, $L_3=300$, $Q=2000$, and $m=1$ 
are chosen. 
One-dimensional condensation occurs in this case.
$T_c = 2.03$ is the three-dimensional critical temperature in Eq. 
(\ref{1dcond-T32}).
In Fig. 4c, $L_1=2$, $L_2=200$, $L_3=200$, $Q=8000$, and $m=0.5$ 
are chosen.
$T_c = 0.98$ is the three-dimensional critical temperature in Eq. 
(\ref{2dcond-T32}).
Two-dimensional condensation can be seen.  
In Fig. 4d, condensation occurs in three-steps.
$L_1=2$, $L_2=100$, $L_3=600$, $Q=4000$, and $m=0.5$ 
are used.
The long dashed line is the chemical potential $\mu$.
$T_c = 0.79$ is the three-dimensional critical temperature in Eq. 
(\ref{3dcond-T32}).
The logarithmic $T$ scale is used in Fig. 4e for the parameters in Fig. 4d.


%
\end{document}